\documentclass{article}

\usepackage{fullpage}

\usepackage{lineno,natbib}
\modulolinenumbers[5]

\usepackage{amsmath}
\usepackage{amsfonts}
\usepackage{bm} 
\usepackage{tabularx,multirow} 
\usepackage{booktabs}
\usepackage{color}
\usepackage{graphicx}
\usepackage{array} 
\usepackage{paralist} 
\usepackage{verbatim} 
\usepackage{subfig} 
\usepackage[utf8]{inputenc} 
\usepackage[affil-it,auth-sc]{authblk}
\usepackage[hidelinks]{hyperref}

\DeclareMathOperator{\divg}{\mbox{div}} 
\DeclareMathOperator{\divgs}{\mbox{div}_s} 
\DeclareMathOperator{\grads}{\nabla_{\!s}} 

\newcommand{\mb}[1]{\mathbf{#1}}

\newcommand{\mc}[1]{\mathcal{#1}} 
\newcommand{\osc}[1]{\mathring{#1}}
\newcommand{\wt}[1]{\widetilde{#1}} 
\newcommand{\wh}[1]{\widehat{#1}} 

\newcommand{\deltanu}{\delta}

\begin{document}
	
\title{Large-strain poroelastic plate theory for polymer gels with applications to swelling-induced morphing of composite plates}
	
\author[1,\footnote{Corresponding author -- e-mail address:\,\texttt{alucanto@sissa.it}.}]{Alessandro Lucantonio}
\author[2]{Giuseppe Tomassetti}
\author[1]{Antonio DeSimone}
\affil[1]{\small SISSA - International School for Advanced Studies, via Bonomea 265, 34136 Trieste - Italy}
\affil[2]{\small Dipartimento di Ingegneria Civile e Ingegneria Informatica, Università degli Studi di Roma ``Tor Vergata'', via Politecnico 1, 00133 Roma - Italy}
	
\date{}

\maketitle
	
\begin{abstract}
We derive a large-strain plate model that allows to describe transient, coupled processes involving elasticity and solvent migration, by performing a dimensional reduction of a three-dimensional poroelastic theory. We apply the model to polymer gel plates, for which a specific kinematic constraint and constitutive relations hold. Finally, we assess the accuracy of the plate model with respect to the parent three-dimensional model through two numerical benchmarks, solved by means of the finite element method. Our results show that the theory offers an efficient computational framework for the study of swelling-induced morphing of composite gel plates.
\\ \medskip \\
\noindent{\bf Keywords:} plates, large strain, polymer gel, swelling.
\end{abstract}

\section{Introduction}
The analysis of plates and shells undergoing large strains has attracted a considerable research effort, especially in computational mechanics \citep{braun_nonlinear_1994,sansour_theory_1995,basar_finite-element_1996,basar_shear_1997,sussman_3D-shell_2013}. So far, attention has been essentially restricted to hyperelastic and viscoelastic materials, including rubber-like materials. On the other hand, in the context of thin elastic structures consisting of polymer gels, where rubber-like elasticity is coupled with the motion of a fluid that causes the material to swell, large-strain poroelastic plate models are needed. The complexity of current theories of polymer gel swelling \citep{Hong2008,Wang2012,AL2013} stems from their non-linear, multiphysics and three-dimensional character and typically demands for computationally intensive numerical solutions \citep{Zhang2009,Bouklas2015,Chester2015}. In partial response to these issues, the formulation of dimensionally-reduced theories, including poroelastic plate theories, 
 will allow to develop efficient computational models, and occasionally to gain insight from analytical or semi-analytical solutions, which are seldom found for three-dimensional models. 

In this spirit but in a different context, a number of plate and shell models has been developed in recent years to describe the mechanics of thin elastic bodies undergoing material growth \citep{Efrati2009,Dervaux2009,Lewicka2010}. These models were mainly motivated by the study of shape formation and mechanical instabilities in natural and biological thin structures \citep{Goriely2005}. Later, these theories have been applied to model shape morphing of responsive gels \citep{Sharon2010,Dias2011} by assimilating swelling to growth. Specifically, in such applications, photolitographic patterning of the cross-linking density of thin gel membranes \citep{Klein2007, Kim2012a, Wu2013} and, more generally, fabrication of composite thin gel structures enable three-dimensional transformations through non-homogeneous or anisotropic swelling. Despite their success in reproducing experimental results, these models do not include any thermodynamic treatment of swelling  and, as a result, they lack the relation between elasticity of the network and degree of swelling. Such relation is usually introduced phenomenologically through experimental calibration. Finally, all the proposed growth-based models lack an evolution structure, since they do not include the balance of solvent mass, and thus only allow to study the equilibrium shapes of swelling plates \citep{Gemmer2013}. Transient swelling processes are fundamental, for instance, in the toughening of homogeneous \citep{Noselli2016} and composite gels \citep{PhysRevLett.115.188105,Lucantonio2016Mecc}.

Motivated by the observations above, here we introduce a large-strain plate theory for polymer gels obtained by consistent dimensional reduction of the three-dimensional, coupled elasticity-solvent migration model established in \cite{AL2013}. Our theory is not restricted to static equilibrium shapes, but it is also suitable for the study of approach to equilibrium and transient phenomena. This work extends the recent results presented in \cite{Lucantonio2015}, where a membrane theory for swelling polymer gels has been introduced. Our deductive strategy of dimensional reduction employs the weak formulation of the three-dimensional governing equations as a tool to derive the local balance laws for the plate, including the balance of forces, moments and solvent mass. The same deductive approach is employed to obtain the dimensionally reduced counterpart of the three-dimensional swelling constraint, which relates the solvent volume fraction to the volume change of the plate. Thermodynamical consistency is guaranteed through the dissipation inequality, which provides the appropriate constitutive equations for the plate theory, upon introducing for the state variables the same representations along the thickness as those used for the virtual fields. We specialize the constitutive equations for polymer gel plates by employing the Flory-Rehner free energy. In the last section,  we study three numerical benchmarks to validate the present plate theory with respect to the parent three-dimensional model. In particular, we show how the proposed plate model can be employed as a computational tool for the shape programming of composite gel plates.

\section{Geometry and kinematics of the plate}
\label{sec:kinematics}
The ambient space of our theory is represented by the three-dimensional Euclidean point space $\mc{E}$, whose associated space of translations is denoted by $\mc{V}$. Unless otherwise stated, free Latin indices range over $\{0,1,2\}$, free Greek indices range over $\{1,2\}$, and the standard summation convention over repeated indices is employed. Explicit dependence on time of the fields of interest is omitted. 

In its reference configuration, the plate occupies a cylindrical region $\mathcal B$ of height $h$ and mid--surface $\mathcal S$. For convenience in later computations we parametrize the mid--surface through a generic coordinate system $(s^1,s^2)$ ranging in a domain $\mathcal M\subset\mathbb R^2$. Denoting by $\osc r(s)$ the point on $\mathcal S$ identified by the coordinates $s=(s^1,s^2)$ and letting $\osc{\mb n}$ denote the unit vector orthogonal to $\mathcal S$, we can write the typical point of the plate as
\begin{align}\label{eq:22}
  \osc p(s,\zeta) = \osc{r}(s) +\zeta \osc{\mb{n}}\,,
\end{align} 
with $\zeta$ ranging in the interval $(-h/2,+h/2)$. We assume that the function $\osc r:\mathcal M\to\mathcal S$ is smooth enough so that the forthcoming differential operations performed on $\osc r$ make sense.

Expression \eqref{eq:22} provides a parametrization in terms of coordinates $(s^1,s^2,\zeta)$ for the reference configuration $\mathcal B$ of the plate. The spatial dependence of a (scalar, vectorial, or tensorial) field $f$ defined on $\mathcal B$ is expressed using this coordinate system, and we write $f_{,\alpha}=\frac {\partial f}{\partial s^\alpha}$ and $f_{,3}=\frac{\partial f}{\partial\zeta}$. In particular, we denote by
\[
\osc{\mb g}_{\alpha}=\osc p_{,\alpha}=\osc r_{,\alpha},\qquad \osc{\mb g}_{3}=\osc{p}_{,3}=\osc{\mb n}
\]
the referential covariant basis.

With this notation, we can write the  area element of the reference base surface as $\mbox{d} a = \osc\Gamma\,\mbox{d}s^1\mbox{d}s^2$, with $\osc\Gamma = \osc{\mb{g}}_1 \times \osc{\mb{g}}_2 \cdot \osc{\mb{n}}$. Moreover, we denote by $\bm{\nu}(s)$ the outward unit normal to $\partial\mc{S}$ at $\osc p(s)$, and by $\ell$ the arc-length parameter of $\partial \mc{S}$. Calculus in curvilinear coordinates yields the formulas
\begin{equation}\label{eq:14}
\nabla f=\grads f+f_{,3}\otimes\osc{\mb n}\,,\qquad \grads f=f_{,\alpha}\otimes\osc{\mb g}^{\alpha}\,,
\end{equation}
(here, the tensor product between a scalar and a vector is the standard multiplication) where the formulas
\begin{align}
\osc{\mb g}^1=\osc{\Gamma}^{-1}{\osc{\mb g}_{2}\times \osc{\mb n}}\,,\qquad \osc{\mb g}^2=\osc{\Gamma}^{-1}{\osc{\mb n}\times\osc{\mb g}_{1}}\,,\qquad \osc{\mb g}^3=\osc{\mb n}\,.
\end{align}
express the standard contravariant basis. 

The configuration of the plate is described, at each time, by the pair of fields $(p(s,\zeta),c(s,\zeta))$, where $p(s,\zeta)$ is the \textit{deformation} of the plate and $c(s,\zeta)$ is the \textit{solvent concentration} per unit reference volume.  Plate theories are usually generated by approximating $p(s,\zeta)$ with an expression involving a finite number of fields that depend on $s$ only. The theory we propose in this paper approximates $p(s,\zeta)$ with the expression
\begin{align}\label{eq:deformation}
 p(s,\zeta) = r(s)+\zeta\left(1+\frac{{\deltanu}(s)}2\zeta\right) \mb{d}(s)
\end{align}
involving three fields: the \emph{placement of the mid--surface} $r:\mathcal M\to\mathcal E$, the \emph{director} $\mb d:\mathcal M\to\mathcal V$, and the \emph{scalar corrector field} $\delta:\mathcal M\to\mathbb R$. 
In what follows, we denote by $\mb a_{\alpha}=r_{,\alpha}$ the covariant basis of the tangent plane to the current configuration of the mid--surface.
According to \eqref{eq:deformation} the deformation gradient can be expressed as
\begin{align}\label{eq:F}
\mb{F} = \nabla p= \mb{g}_\alpha\otimes\osc{\mb{g}}^\alpha + \mb{g}_3\otimes\osc{\mb{n}}\,,
\end{align}
where
\begin{equation}\label{eq:19}
\mb g_\alpha=\mb a_{\alpha}+\zeta\left(1+\frac \delta 2\zeta\right)\mb d_{,\alpha}+\frac {\delta_{,\alpha}}2\mb d\,,\qquad \mb g_3=\left(1+\zeta\delta\right)\mb d\,,
\end{equation}
is the current covariant basis. The vector $\mb g_3(s,\zeta)$ represents the image under the deformation $p(s,\zeta)$ of an infinitesimal material fiber initially placed at $\osc p(s,\zeta)$ and parallel to $\osc{\mb n}$. As is apparent from \eqref{eq:19} the fiber becomes parallel to $\mb d$ and undergoes a non--uniform stretch:
\begin{align}
\lambda_3(s,\zeta)=\sqrt{\mb F\osc{\mb n}\cdot\mb F\osc{\mb n}}=\sqrt{\mb g_3(s,\zeta)\cdot\mb g_3(s,\zeta)}=|(1+\delta(s)\zeta)\mb d(s)|,
\end{align}
which does not vanish within the interval $[-h/2,h/2]$ provided that $|\delta(s)|<2/h$. The introduction of the scalar field ${\deltanu}$ in the kinematics of the plate results into a 7-parameter (the other six parameters are the three components of the displacement $\bm u(s)=r(s)-\osc r(s)$ of the base surface and the three components of the deformed director $\mb d(s)$) theory that has been shown to be effective in avoiding locking problems \citep{braun_nonlinear_1994} in compressible materials, without performing any manipulation of the three-dimensional constitutive equations before dimensional reduction. Typically, in models using a plane-strain (constant thickness) kinematics, locking is avoided by enforcing the plane-stress hypothesis at the three-dimensional level, which is not compatible with plane-strain, but leads nevertheless to a theory that provides accurate results in many cases.  

For the sake of calculation, we shall oftentimes find it is convenient to rewrite the deformation as
 \begin{align}
 \label{eq:altdeform}
 p(s,\zeta)=r(s)+\zeta \mb{d}_1(s)+ \frac{\zeta^2}{2} \mb{d}_{2}(s)\,, \qquad \mb d_1\equiv \mb d\,, \qquad \mb d_2\equiv {\deltanu} \mb d\,.    
 \end{align}
 Then, letting $\wh{\mb{F}}=\grads r=\mb a_{\alpha}\otimes\osc{\mb g}^{\alpha}$, we can rewrite \eqref{eq:F} as   
 \begin{align}
 \label{eq:defgradshell}
 \mb{F} = \left(\wh{\mb{F}}+\zeta\grads{\mb{d}_1} + \frac{\zeta^2}{2} \grads{\mb{d}_2}\right)+ (\mb{d}_1+\zeta\mb{d}_2) \otimes \osc{\mb{n}}\,.
 \end{align}
 As we will see in Section~\ref{sec:consteq}, this two-director representation of $\mb{F}$ is particularly suitable to express the reduced constitutive equations.

Further, we take the concentration $c$ to be linearly dependent on the thickness coordinate:
\begin{align}
\label{eq:conc}
c(s,\zeta) = c_{0}(s)+\zeta c_1(s)\,.
\end{align}
Importantly, we observe that, in principle, neither \eqref{eq:deformation} nor  \eqref{eq:conc} are necessary for the determination of the balance equations of the plate theory from the principle of virtual power, since for that aim only the representations of the virtual fields are involved. Thus, the use of  \eqref{eq:deformation}, \eqref{eq:conc} is effectively restricted to performing explicit integrations along the thickness in the derivation of reduced constitutive equations (see Section~\ref{sec:consteq}).

\section{Balance equations}
\label{sec:balanceeq}

\subsection{Balance of forces and moments}
Following \cite{AntmanBook,DiCarlo2001}, we derive the balance of forces and moments for the plate-like body $\mc{B}$ starting from the principle of virtual power
\begin{align}
\label{eq:virtpowerbal}
\mc{I}^h_{\rm m}(\mc{M},\wt{\mb{v}}) = \mc{W}^h_{\rm m}(\mc{M},\wt{\mb{v}})\,,
\end{align}
which states the equality between the \textit{virtual} expenditures of \textit{internal mechanical power} 
\begin{align}
\label{eq:intvirtpower}
\mc{I}^h_{\rm m}(\mc{M},\wt{\mb{v}}) = \int_{\mc{M}}\int_{-h/2}^{h/2}{\mb{S}\cdot\nabla\wt{\mb{v}}\,\mbox{d}\zeta\mbox{d}a}
\end{align}
and \textit{external mechanical power} 
\begin{align}
\mc{W}^h_{\rm m}(\mc{M},\wt{\mb{v}}) = \int_{\mc{M}}\int_{-h/2}^{h/2}{\mb{f}\cdot\wt{\mb{v}}}\,\mbox{d}\zeta\mbox{d}a +  
\int_{\mc{M}}\left(\mb{t}\cdot\wt{\mb{v}}\right)\Big|_{\pm h/2}\,\mbox{d}a + 
\int_{\partial\mc{M}}\int_{-h/2}^{h/2}{\mb{t}\cdot\wt{\mb{v}}}\,\mbox{d}\ell\,,
\end{align}
for any  virtual velocity field $\widetilde{\mb{v}}$. Here, $\mb{S}$ is the Piola-Kirchhoff stress and the system of external forces is represented by the body force $\mb{f}$ acting on $\mc{B}$ and the surface traction $\mb{t}$ acting on $\partial\mc{B}$.
The crucial point when using \eqref{eq:virtpowerbal} is to restrict test velocities to those compatible with the kinematic constraint that generates the theory. Here, consistent with the expression \eqref{eq:altdeform} of the deformation of the plate, we assume that, for $s$ fixed, the virtual velocity field is a quadratic polynomial of $\zeta$:
\begin{align}\label{eq:4}
\wt{\mb{v}}(s,\zeta) = \wt{\mb{v}}_{0}(s)+\zeta {\wt{\mb{v}}}_{1}(s)+\frac {\zeta^2}2 {\wt{\mb{v}}}_{2}(s)\,,
\end{align}
so that the internal virtual power expended is
\begin{align}
\label{eq:intpow}
\begin{split}
\int_{\mc{M}}\int_{-h/2}^{h/2}{\left[\bm{S}\cdot \left(\grads{\wt{\mb{v}}_{0}} + \zeta \grads{{\wt{\mb{v}}}_{1}} + \frac{\zeta^2}2 \grads{{\wt{\mb{v}}}_{2}}\right)+\bm{s}\cdot({\wt{\mb{v}}}_{1}+\zeta{\wt{\mb{v}}}_{2})\right]\,\mbox{d}\zeta\mbox{d}a}\,,
\end{split}
\end{align}
where $\bm{S}=\mathbf S\osc{\mathbf P}$, $\bm{s}=\mathbf S\osc{\mb n}$ with $\osc{\mb P}=\osc{\mb g}_\alpha\otimes\osc{\mb g}^{\alpha}$ the in-plane projector.
Introducing the \emph{stress resultants}\footnote{Following standard convention we set $0!=1$.}
\begin{align}\label{eq:40}
\bm{S}^{(i)}(s)=\int_{-h/2}^{h/2}\frac{\zeta^{i}}{i!}\bm{S}(s,\zeta)\,\mbox{d}\zeta\,, && \bm{s}^{(i)}(s)=\int_{-h/2}^{h/2}\frac{\zeta^{(i)}}{i!}{\mb s}(s,\zeta)\,\mbox{d}\zeta\,,
\end{align}
we can recast  \eqref{eq:intpow} as
\begin{align}
\label{eq:intpow2}
{\mc I}_{\rm m}(\mc{M},\wt{\mb{v}}_i)=\int_{\mc{M}}\left(\bm{S}^{(i)}\cdot\grads\wt{\mb v}_{i}+{{\bm s}}^{(0)}\cdot\wt{\mb v}_{1}+{{\bm s}}^{(1)}\cdot\wt{\mathbf v}_{ 2}\right)\mbox{d}a\,.
\end{align}
Likewise, introducing the \emph{bulk force resultants}
\begin{align}
\mb{f}^{(i)}(s)= \int_{-h/2}^{h/2}{\frac{\zeta^{(i)}}{i!}\mb{f}(s,\zeta)\,\mbox{d}\zeta} + \left(\frac{\zeta^{(i)}}{i!}\mb{t}(s,\zeta)\right)\Big|_{\zeta=-h/2}+ \left(\frac{\zeta^{(i)}}{i!}\mb{t}(s,\zeta)\right)\Big|_{\zeta=h/2}
\end{align}
defined for $s\in\mc{M}$, and the \emph{boundary force resultants}
\begin{align}
\mb{t}^{(i)}(s) = \int_{-h/2}^{h/2}{\frac{\zeta^{(i)}}{i!}\mb{t}(s,\zeta)\,\mbox{d}\zeta}
\end{align}
defined for $s\in\partial\mathcal M$, we can write the external power as
\begin{align}
\label{eq:extpow2}
{\mc W}_{\rm m}(\mc{M},\wt{\mb{v}}_i)= \int_{\mc{M}}\left(\mb{f}^{(i)} \cdot \wt{\mb{v}}_{i}\right)\mbox{d}a +  \int_{\partial\mc{M}}\left(\mb{t}^{(i)} \cdot \wt{\mb{v}}_{i}\right)\mbox{d}\ell\,.
\end{align}

When deriving balance equations in strong form by exploiting the arbitrariness of the virtual fields, some care is required, because the virtual velocities $\wt{\mathbf v}_{1}$ and $\wt{\mathbf v}_{2}$ are not independent. Indeed, it holds $\wt{\mb v}_{2}=\wt{\deltanu}\mb d+{\deltanu}\wt{\mb v}_{1}$, since $\mathbf d_2={\deltanu}\mathbf d$. Substitution of this relation into the virtual--power functionals \eqref{eq:intpow2} and \eqref{eq:extpow2}, yields, for the internal power
\begin{equation}\label{eq:35}
{\mc I}'_{\rm m}(\mc{M},\wt{\mb{v}}_0,\wt{\mb{v}}_1,\wt{\deltanu})=
\int_{\mc{M}}\left(\mb{N}\cdot\grads\wt{\mb v}_{0}+\mb{M}\cdot\grads\wt{\mb v}_{ 1}+\mb{q}\cdot\wt{\mb v}_{1}+\bm\tau\cdot\grads\wt\deltanu+\tau\wt\deltanu\right)\mbox{d}a\,,
\end{equation}
where 
\begin{equation}\label{eq:18}
\begin{aligned}
&\mb{N}=\bm{S}^{(0)}\,, && \mb{M}=\bm{S}^{(1)}+{\deltanu}{\bm S}^{(2)}\,, && \mb{q}=
{{\bm s}}^{(0)}+{\deltanu}{\bm s}^{(1)}+\bm S^{(2)}\grads{\deltanu}\,,\\
&{\bm \tau}={{\bm S}^{(2)}}^{\rm T}{\mb d}\,, && \tau={\bm s}^{(1)}\cdot\mathbf d+\bm S^{(2)}\cdot\grads\mb{d}\,, &&
\end{aligned}
\end{equation}
and, for the external power
\begin{align}
{\mc W}'_{\rm m}(\mc{M},\wt{\mb{v}}_0,\wt{\mb{v}}_1,\wt{\deltanu})=\int_{\mc{M}}\left(\mb f^{(0)}\cdot\wt{\mb v}_{0}+\overline{\mb f}^{(1)}\cdot\wt{\mb v}_{1}+ f_d^{(2)}\wt\deltanu\right)\mbox{d}a+\int_{\partial\mc{M}}\left(
\mb t^{(0)}\cdot\wt{\mb v}_{0}+\overline{\mb t}^{(1)}\cdot\wt{\mb v}_{1}+ t^{(2)}_d\wt\deltanu
\right)\mbox{d}\ell\,,
\end{align}
where
\begin{align}
\overline{\mb f}^{(1)}=\mb f^{(1)}+{\deltanu}\mb f^{(2)}\,, && f_d^{(2)}=\mb f^{(2)}\cdot\mb d\,, && \overline{\mb t}^{(1)}=\mb t^{(1)}+{\deltanu}\mb t^{(2)}\,, && t_d^{(2)}=\mb{t}^{(2)}\cdot\mb d\,.
 \end{align}
For simplicity, we assume that the boundary $\partial\mathcal M$ is partitioned into disjoint parts $\partial_{\rm m}^0\mathcal M$ and $\partial_{\rm m}^1\mathcal M$, with natural (essential) boundary conditions being prescribed on the former (latter). Then, from the arbitrariness of $(\wt{\mb{v}}_{0},\wt{\mb{v}}_1,\wt{\deltanu})$ in the virtual--power balance
\begin{align}
\label{eq:redvirtpow}
\mc I'_{\rm m}(\mc{M},\wt{\mb{v}}_{0},\wt{\mb{v}}_1,\wt{\deltanu})=\mc W'_{\rm c}(\mc{M},\wt{\mb{v}}_{0},\wt{\mb{v}}_1,\wt{\deltanu})
\end{align}
we deduce the balance equations and the corresponding natural boundary conditions, namely, 
\begin{subequations}\label{eq:43}
\begin{align}
&\divgs\mathbf N+\mathbf f^{(0)}=\mathbf 0\,, && \mbox{in}\ \mc{M}\,,\\
&\divgs\mb{M}-\mb q+\overline{\mb{f}}^{(1)}=\mathbf 0\,, && \mbox{in}\ \mc{M}\,,\label{eq:balamoments}\\
&\divgs\bm\tau-\tau+f_d^{(2)}=0\,, && \mbox{in}\ \mc{M}\,, \label{eq:baladirforces2}
\end{align}
\end{subequations}
and
\begin{subequations}\label{eq:16}
\begin{align}
&\mb N\bm\nu=\mb t^{(0)}\,, && \mbox{on}\ \partial^0_{\rm m}\mc{M}\,,\\
&\mb M\bm\nu=\overline{\mb t}^{(1)}\,, && \mbox{on}\ \partial^0_{\rm m}\mc{M}\,,\\
&\bm\tau\cdot\bm\nu=t_d^{(2)}\,, && \mbox{on}\ \partial^0_{\rm m}\mc{M}\,.
\end{align}
\end{subequations}
It is easily checked that the symmetry condition $\mb S\mb F^{\rm T}\in{\rm Sym}$ entails
\begin{equation}\label{eq:25}
\mb{N}\wh{\mb{F}}^{\rm T}+\mb{M}\grads \mb{d}^{\rm T}+\mb{q}\otimes\mb{d}\in\mbox{Sym}.
\end{equation}
The tensorial identity \eqref{eq:25} is equivalent to
\begin{equation}
\label{eq:symstress}
\mb n^\alpha\times\mb a_\alpha+\mb m^\alpha\times\mb{d}_{,\alpha}+\mb q\times\mb d = \mb{0}\,,
\end{equation}
where $\mb n^\alpha=\mb N\osc{\mb g}^{\alpha}$  and   $\mb m^\alpha=\mb M\osc{\mb g}^{\alpha}$. This relation can also be obtained by imposing invariance of the internal power expended within every part of the plate under superposed rigid velocity fields. In view of  \eqref{eq:symstress}, by taking the cross and scalar products of  \eqref{eq:balamoments} with $\mb{d}$ we obtain, respectively,
\begin{subequations}
\label{eq:balamomclass}
\begin{align}
&\divgs(\mb{D}\mb{M})-\mb n^\alpha\times \mb a_\alpha+{\mb d}\times\overline{\mb{f}}^{(1)}=\mb 0\,, \\
&\divgs(\mb M^{\rm T}\mb d)-\mb M\cdot\grads\mb d-\mb d\cdot\mb q+\mb d\cdot\overline{\mb f}^{(1)}=0\,,
\end{align}
\end{subequations} 
where $\mb{D}$ is the skew tensor associated to $\mb{d}$. From  \eqref{eq:balamomclass}, we can recover the classical forms of the balances of torques and director forces for a 6-parameter plate by setting ${\deltanu}=0$, see \cite{AntmanBook,DiCarlo2001}. For a 7-parameter plate, these equations are supplemented by the second-order balance of director forces \eqref{eq:baladirforces2}.

\subsection{Balance of solvent mass}
In the applications consider here, a flux of solvent is prescribed over the lateral surface of the plate, including the top and bottom faces. The strong form of the balance of solvent mass reads 
\begin{subequations}
\label{eq:balasol3d}
\begin{align}
&\dot c+\divg\mb h=0\,,     &&\mbox{in }\mc B\,,\\  
&-\mb h\cdot\bm\nu=\beta\,, &&\mbox{on }\partial\mc M\times(-h/2,+h/2)\,,
\end{align}
\end{subequations}
where $\mb h$ is the \emph{referential solvent flux}, and $\beta$ is a surface supply of solvent.

With a view towards obtaining a reduced theory, we replace the pointwise statements \eqref{eq:balasol3d} with their weak form. Inspired by \cite{Duda2010}, we interpret such weak form as a version of the virtual-power principle whereby chemical potential is a test field that enters along with its gradient in the power expenditure. Specifically, we prescribe the following representations for \textit{virtual} expenditures of \textit{internal chemical power}\begin{align}
\label{eq:intchempow}
\mc{I}^h_{\rm c}(\mc{M},\wt{\mu})=\int_{\mc{M}}\int_{-h/2}^{h/2} (\dot{c}\wt{\mu}-\mb{h}\cdot\nabla\wt{\mu})\,\mbox{d}\zeta \mbox{d}a\,, 
\end{align}
and \emph{external chemical power}
\begin{equation}
\label{eq:extchempow}
\mc{W}^h_{\rm c}(
\mc{M},\wt{\mu})=\int_{\partial \mc{M}}\int_{-h/2}^{h/2}\beta\wt{\mu}\,\mbox{d}\zeta\mbox{d}\ell+\int_{\mathcal M\times\left\{+\frac h 2,-\frac h 2\right\}}\beta\widetilde\mu\,\mbox{d}a\,.
\end{equation}
Then, the three-dimensional pointwise balance equations \eqref{eq:balasol3d} are recovered on imposing that the external and internal chemical powers be balanced 
\begin{equation}
\label{eq:virtchembal}
\mc{I}^h_{\rm c}(\mc{M},\wt{\mu})=\mc{W}^h_{\rm c}( 
\mc{M},\wt{\mu})
\end{equation}
for any \textit{virtual chemical potential} $\wt{\mu}$. 

To arrive at a system of reduced equations, we enforce the principle of virtual power \eqref{eq:virtchembal} on the class of  virtual chemical potentials that depend linearly on $\zeta$:
\begin{equation}
\label{eq:virtchempot}
\wt{\mu}(s,\zeta) = \wt{\mu}_{0}(s) + \zeta \wt{\mu}_{1}(s)\,,
\end{equation}
consistent with the representation \eqref{eq:conc} for the work-conjugate field $c$.
Granted  \eqref{eq:virtchempot} and using  \eqref{eq:14} with $\phi=\wt{\mu}$, we can write the internal chemical power  as:
\begin{equation}
\label{eq:2dintvirtchempow}
\mc{I}_c(\mc{M},\wt{\mu}_0,\wt{\mu}_1)=\int_{\mc{M}}\left(\dot c^{(0)}\wt{\mu}_{0}+ \dot c^{(1)}\wt{\mu}_{ 1}-\bm{h}^{(0)}\cdot\grads{\wt{\mu}_{0}}-h_3^{(0)}\wt{\mu}_{1}-\bm{h}^{(1)}\cdot\grads{\wt{\mu}_{1}}\right)\mbox{d}a\,,
\end{equation}
where we have introduced the moments of the concentration
\begin{equation}
c^{(i)}(s)=\int_{-h/2}^{h/2}\frac{\zeta^{i}}{i!}c(s,\zeta)\,\mbox{d}\zeta\,,
\end{equation}
and of the solvent flux
\begin{subequations}
\label{eq:momflux}
\begin{align}
 &\bm{h}^{(i)}(s)=\int_{-h/2}^{h/2}\frac{\zeta^{i}}{i!}\bm{h}(s,\zeta)\,\mbox{d}\zeta\,, \quad \bm h=\osc{\mb P}\mb{h}\,,\\
 &h_3^{(i)}(s)=\int_{-h/2}^{h/2}\frac{\zeta^{i}}{i!}h_3(s,\zeta)\,\mbox{d}\zeta\,, \quad h_3=\mb h\cdot\osc{\mb{a}}_3\,.
\end{align}
\end{subequations}
Likewise, the external power becomes
\begin{equation}
\label{eq:extchempot}
\mc{W}_{\rm c}(\mc{M},\wt{\mu}_0,\wt{\mu}_1)=\int_{\partial\mc{M}}(\beta^{(0)}\wt{\mu}_{0}+\beta^{(1)}\wt{\mu}_{1})\mbox{d}\ell+\int_{\mathcal M}(\overline{\beta}^{(0)}\wt\mu_0+\overline\beta^{(1)}\wt\mu_1)\,\mbox{d}a\,.
\end{equation}
where
\begin{subequations}
  \begin{align}
     \beta^{(i)}(s)&=\int_{-h/2}^{h/2}\frac{\zeta^{i}}{i!}\beta(s,\zeta)\,\mbox{d}\zeta\, ,\qquad s\in\partial\mathcal M\,,\\
     \overline\beta^{(i)}(s)&=\sum_{\zeta=\pm\frac h 2}\frac{\zeta^{i}}{i!}\beta(s,\zeta)\,\mbox{d}\zeta\,,\qquad s\in\mathcal M.
\end{align}
\end{subequations}
With these definitions the principle of virtual power \eqref{eq:virtchembal} may be written as
\begin{align}
\label{eq:redvirtchempow}
\mc{I}_{\rm c}(\mc{M},\wt{\mu}_0,\wt{\mu}_1) = \mc{W}_{\rm c}(\mc{M},\wt{\mu}_0,\wt{\mu}_1)
\end{align}
for any $\wt{\mu}_0$ and $\wt{\mu}_1$.

We partition the boundary of $\mathcal M$ into parts $\partial_{\rm c}^0\mathcal M$ and $\partial_{\rm c}^1\mathcal M$, and we impose essential boundary conditions for the chemical potential on $\partial_{\rm c}^1\mathcal M$. On exploiting the arbitrariness of the virtual chemical potential, \textit{i.e.} of the fields $\wt{\mu}_0$ and $\wt{\mu}_1$, in $\mathcal M\cup\partial_{\rm c}^0\mathcal M$, we obtain the balance equations
\begin{subequations}\label{eq:redbalasol}
\begin{align}
&\dot c^{(0)}+\divgs{\bm{h}^{(0)}}=\overline \beta^{(0)}\,, && \mbox{in}\ \mc{M}\,, \\
&\dot c^{(1)}+\divgs{\bm{h}^{(1)}}-h_3^{(0)}=\overline \beta^{(1)}\,, && \mbox{in}\ \mc{M}\,,
\end{align}
\end{subequations}
and the natural boundary conditions
\begin{subequations}\label{eq:21}
\begin{align}
&-\bm{h}^{(0)}\cdot\bm\nu = \beta^{(0)}\,, && \mbox{on}\ \partial_{\rm c}^0\mc{M}\,, \\
&-\bm{h}^{(1)}\cdot\bm\nu= \beta^{(1)}\,, && \mbox{on}\ \partial_{\rm c}^0\mc{M}\,.  
\end{align}
\end{subequations}
Notice that the above equations do not give explicitly the concentration field. However, if we assume that the concentration field has the representation \eqref{eq:conc}, then we can obtain from \eqref{eq:redbalasol} a set of evolution equations for $c_{0}$ and $c_{1}$. To this aim, it suffices to make use of the following identities:
\begin{subequations}
\label{eq:17}
\begin{align}
&c^{(0)} = h c_{0}\,,\qquad c^{(1)} = \frac {h^3}{12}c_{1}\,.
\end{align}
\end{subequations}
Then, the balance equations \eqref{eq:redbalasol} become
\begin{subequations}\label{eq:44}
	\begin{align}
	&h\dot c_{0}+\divgs{\bm{h}^{(0)}}=\overline \beta^{(0)}\,, && \mbox{in}\ \mc{M}\,, \\
	&\frac{h^3}{12}\dot c_{1}+\divgs{\bm{h}^{(1)}}-h_3^{(0)}=\overline \beta^{(1)}\,, && \mbox{in}\ \mc{M}\,.
	\end{align}
\end{subequations}

\noindent{\bf Remark: essential boundary conditions for chemical potential.} In the considerations leading to \eqref{eq:44}, essential boundary conditions for chemical potential have been imposed only on the lateral mantle $\partial\mathcal M\times\left(-\frac h 2,+\frac h 2\right)$ of the plate. This restriction can be easily removed, and one can impose essential conditions on parts of the top and bottom face of the plate, to be handled by making use of Lagrange multipliers. To be specific: let the boundary values of the chemical potential field be assigned values $\mu^-$ and $\mu^+$ on parts of the top and bottom faces, which we denote by $\mathcal M^+\times\left\{\frac h 2\right\}$ and  $\mathcal M^-\times\left\{\frac h 2\right\}$,  with $\mathcal M^+\subseteq\mathcal M$ and $\mathcal M^-\subseteq\mathcal M$. Then, denoting by $\chi^+:\mathcal M\to\{0,1\}$ and $\chi^-:\mathcal M\to\{0,1\}$ the characteristic functions of $\mathcal M^+$ and $\mathcal M^-$, respectively, we add the following reactive term 
\[
\mathcal W_{\rm c}^r(\mathcal M,\wt\mu_0,\wt\mu_1)=\int_{\mathcal M}\Big(\chi^+r^+\big(\wt\mu_0+\frac h 2\wt \mu_1\big)+\chi^-r^-\big(\wt\mu_0+\frac h 2\wt \mu_1\big) \Big)\,\mbox{d}a
\]
to the external power \eqref{eq:extchempot}. Then, the additional reactive contributions $\chi^+ r^++\chi^- r^-$ and $\frac h 2(\chi^+r^+-\chi^-r^-)$ appear on the right--hand sides of the first and second equation in \eqref{eq:44}, respectively. 

\section{Swelling constraint}
For a poroelastic medium undergoing large strain, the assumption of incompressibility of both the solid and the solvent implies that the deformation gradient $\mb F$ and the concentration $c$ obey the \emph{incompressibility constraint} \citep{Hong2008,AL2013}
\begin{align}
\label{eq:volumeconstr}
&J = \det \mb{F} = 1 + \Omega (c-c_\star)\,,
\end{align}
where $c_\star$ is the homogeneous solvent concentration in the reference state. Since the deformation is a polynomial of degree 2 with respect to $\zeta$, its determinant is a polynomial of degree 6 with respect to the same variable. Now, recalling from \eqref{eq:conc} that concentration is a first--order polynomial in $\zeta$, the constraint \eqref{eq:volumeconstr} cannot be satisfied pointwise.

This state of matters leads us to \emph{relax} constraint \eqref{eq:volumeconstr}, by replacing it with a weak constraint
\begin{align}
&\int_{\mc{M}}\int_{-h/2}^{h/2}{[J-1-\Omega(c-c_{\star})]\tilde{p}\,\mbox{d}\zeta\mbox{d}a} = 0\,,
\end{align}
where the \emph{virtual pressure} $\tilde{p}$ has the representation $\tilde p(s,\zeta)=\tilde{p}_{0}(s)+\zeta\tilde{p}_{1}(s)$.
By enforcing the weak form of the incompressibility constraint with degree-1 virtual pressures we are going to deduce, for each point $s$ of the base surface, a set of two equations relating the deformation with the fields $c_{0}$ and $c_{1}$. 

From the identity $J=(\mb F\osc{\mb g}_1\times\mb F\osc{\mb g}_2\cdot\mb F\osc{\mb n})/(\osc{\mb g}_1\times\osc{\mb g}_2\cdot\osc{\mb n})$ and from  \eqref{eq:F} we find $J={\mb g}_1\times{\mb g}_2\cdot{\mb g}_3/\osc\Gamma$. Thus, using \eqref{eq:19} we compute:
\begin{align}
\begin{split}
J &= \left(\mb{r}_{0} + \zeta \mb{r}_1 +\frac{\zeta^2}{2}\mb{r}_2\right)\cdot(1+\zeta\deltanu)\mb{d} + o(\zeta^2)\,,
\end{split}
\end{align}
with
\begin{align}
\label{eq:qs}
&\mb{r}_{0} = \frac{\mb{a}_1 \times \mb{a}_2}{\osc\Gamma}\,, \qquad \mb{r}_1 = \frac{\mb{d}_{,1} \times \mb{a}_{2} - \mb{d}_{,2}\times \mb{a}_1}{\osc\Gamma}\,, \qquad \mb{r}_2 = \deltanu\mb{r}_1 + \frac{(\deltanu_{,2}\mb{a}_1 - \deltanu_{,1}\mb{a}_2)\times\mb{d} + 2\mb{d}_{,1}\times\mb{d}_{,2}}{\osc\Gamma}\,.
\end{align}

Then, we rely on the arbitrariness of $\tilde{p}_{0}(s)$  and $\tilde{p}_{1}(s)$ to obtain the pair of equations holding pointwise in $\mc{M}$:
\begin{subequations}
\label{eq:redswellconstrplate}
	\begin{align}
	1+\Omega(c_{0}-c_{\star}) &= \mb{r}_{0}\cdot\mb d + \frac{h^2}{24}(2\deltanu\mb{r}_1 + \mb{r}_2)\cdot\mb{d}\,, \\
	\Omega c_{1}&=\frac{h^3}{12}(\deltanu\mb{r}_0+\mb{r}_1)\cdot\mb{d}\,. 
	\end{align}
\end{subequations} 

\section{Thermodynamics and constitutive equations}
\label{sec:consteq}
In this section we obtain  constitutive equations for the stress resultants and the flux resultants, starting from those that govern the large-strain behavior of three dimensional poroelastic bodies with incompressible constituents. We first deduce general relations; then, we consider the specialization of these relations to  composite polymer gel plates.

\subsection{General relations}
 We begin by assuming that the free energy per unit referential volume obeys the constitutive equation
\begin{equation}\label{eq:7}
\psi=\widehat\psi(\mathbf F,c).
\end{equation}
For $\mathcal P\subset\mathcal M$ we write the dissipation inequality as
\begin{align}\label{eq:dissipineqint}
\begin{split}
&\int_{\mc{P}}\int_{-h/2}^{h/2}{\left[\dot{\psi}-p\left(\mb{F}^\star\cdot\dot{\mb{F}}-\Omega\dot{c}\right)\right]\mbox{d}\zeta\mbox{d}a} \leq \mc{I}^h_{\rm m}(\mc{P},\mb{v})+\mc{I}^h_{\rm c}(\mc{P},\mu),
\end{split}
\end{align}
where $\mc{I}^h_{\rm m}(\mc{P},\mb{v})$ and $\mc{I}^h_{\rm c}(\mc{P},\mu)$ are, respectively, the mechanical and the chemical powers given in  \eqref{eq:intvirtpower} and \eqref{eq:intchempow} expended within a part $\mc P\times(-h/2,+h/2)$ on the actual velocity $\mb{v} = \dot{p}$ and chemical potential. Here, $\mb F^\star$ is the cofactor of $\mb F$:
\begin{align}
\label{eq:cofactor}
\mb F^\star=J\mb g^i\otimes\osc{\mb g}_i,
\end{align}
where 
\begin{align}
\mb{g}^1 = \frac{\mb{g}_2 \times \mb{g}_3}{\Gamma}\,, \qquad \mb{g}^2 = \frac{\mb{g}_{ 3} \times \mb{g}_1}{\Gamma}\,, \qquad \mb{g}^3 = \frac{\mb{g}_{ 1} \times \mb{g}_2}{\Gamma}\,,\qquad\mbox{with}\qquad \Gamma=\mb{g}_{ 1} \times \mb{g}_2 \cdot \mb{g}_3
\end{align}
are the contravariant basis vectors in the current configuration. 

We consider evolution processes such that the deformation and the concentration obey the restrictions \eqref{eq:deformation} and \eqref{eq:conc}. For any such process, the velocity and the swelling rate (the time derivative of the concentration) have the form
\begin{equation}\label{eq:5}
\mb v(s,\zeta)=\mathbf v_{0}(s)+\zeta\mathbf v_{1}(s)+\frac {\zeta^{2}}2\mathbf v_{2}(s)\qquad \mbox{and}\qquad 
\dot c(s,\zeta)=\dot c_{0}(s)+\zeta \dot c_{1}(s).
\end{equation}
From \eqref{eq:5} and from the constitutive equation \eqref{eq:7} we obtain
\begin{align}
\label{eq:9}
\int_{-h/2}^{h/2}\dot{\psi}\,\mbox{d}\zeta\mbox{d}a=(\partial_{\mb F}\widehat\psi)^{( i)}\cdot\grads\mb v_{i}+(\partial_{\mb F}\widehat\psi)^{(0)}[\osc{\mb n}]\cdot\mb v_{ 1}+(\partial_{\mb F}\widehat\psi)^{( 1)}[\osc{\mb n}]\cdot\mb v_{2}+(\partial_c\widehat\psi)^{(0)}\dot c_{ 0}+(\partial_c\widehat\psi)^{(1)}\dot c_{ 1},
\end{align}
where
\begin{align}
(\partial_{\mb F}\wh\psi)^{(i)}=\left(\int_{-h/2}^{h/2}\frac{\zeta^{i}}{i !}\partial_{\mb F}\wh \psi(\mb F,c)\,\mbox{d}\zeta\right)\osc{\mb P}\qquad\textrm{and}\qquad(\partial_{c}\widehat\psi)^{( i)}=\int_{-h/2}^{h/2}\frac{\zeta^{i}}{i !}\partial_{c}\wh \psi(\mb F,c)\,\mbox{d}\zeta.
\end{align}
Thanks to the identity \eqref{eq:cofactor}, we have the following representation for the pressure power:
\begin{align}
\label{eq:2}
\begin{split}
\int_{-h/2}^{h/2}{p\mb{F}^\star\cdot\dot{\mb{F}}\,\mbox{d}\zeta} &=\bm{S}_p^{(i)}\cdot\grads{\mathbf v}_{ i}+{{\bm s}}_p^{(\rm 0)}\cdot{\mathbf v}_{1}+{{\bm s}}_p^{(1)}\cdot{\mathbf v}_{2},
\end{split}
\end{align}
where
\begin{align}
\label{eq:reactivestress}
&\bm{S}_p^{(i)} = \left(\int_{-h/2}^{h/2}{\frac{\zeta^ i}{i!}p  J \mb{g}^\alpha\,\mbox{d}\zeta}\right)\otimes\osc{\mb{g}}_\alpha \qquad\textrm{and}\qquad \mb{s}_{p}^{(i)} = \int_{-h/2}^{h/2}\frac{\zeta^ i}{i!}p J \mb{g}^3\,\mbox{d}\zeta.
\end{align}

Finally, consistent with the representation of the virtual chemical potential we have selected in \eqref{eq:virtchempot} to enforce mass balance, we assume that the chemical potential has the form
\begin{equation}\label{eq:chempot}
\mu(s,\zeta)=\mu_{ 0}(s)+\zeta\mu_{ 1}(s)\,.
\end{equation}
Then, with calculations analogous to those leading to \eqref{eq:intpow2} and \eqref{eq:2dintvirtchempow}, the internal powers in  \eqref{eq:dissipineqint} may be recast as $\mc{I}_{\rm m}(\mc{P},\mb{v}_i)$ and $\mc{I}_{\rm c}(\mc{P},\mu_0,\mu_1)$, respectively, where the virtual fields in  \eqref{eq:intpow2} and \eqref{eq:2dintvirtchempow} are replaced by the actual fields $\mb{v}_i$ and $\mu_i$.

On account of the reduced representations of the internal powers and the expressions \eqref{eq:9} and \eqref{eq:2} for the time rate of the free energy and the reactive power, the dissipation inequality \eqref{eq:dissipineqint} reads
\begin{align}\label{eq:3}
\begin{split}
&\int_{\mc{P}}\left[\left((\partial_{\mb F}\widehat\psi)^{(i)}+\bm{S}^{(i)}_p\right)\cdot\grads\mb v_{ i}+\left((\partial_{\mb F}\widehat\psi)^{(0)}[\osc{\mb n}]-\bm s^{(0)}_p\right)\cdot\mb v_{ 1}+\left((\partial_{\mb F}\widehat\psi)^{(1)}[\osc{\mb n}] -\bm s^{(1)}_p\right)\cdot\mb v_{ 2}\right]\mbox{d}a\\
&\quad+\int_{\mc{P}}\left[\left((\partial_c\widehat\psi)^{(0)}-p^{(0)}\right)\dot c_{ 0}+\left((\partial_c\widehat\psi)^{(1)}-p^{(1)}\right)\dot c_{1}\right]\mbox{d}a \\ 
&\leq \quad\int_{\mc{P}}\left(\bm{S}^{(i)}\cdot\grads{\mb v}_{i}+{{\bm s}}^{(0)}\cdot{\mb v}_{1}+{{\bm s}}^{(1)}\cdot{\mb v}_{2}\right)\mbox{d}a\\
&\quad +\int_{\mc{P}}\left(\mu^{(0)}\dot{c}_{0}+\mu^{(1)}\dot{c}_{1}-\bm{h}^{(0)}\cdot\grads \mu_{0}-h_3^{(0)}{\mu}_{1}-\bm{h}^{(1)}\cdot\grads{{\mu}_{1}}\right)\mbox{d}a.
\end{split}
\end{align}
Consistent with the requirement that \eqref{eq:3} holds for every choice of the velocities \eqref{eq:5} we prescribe
\begin{subequations}\label{eq:8}
\begin{align}
&\bm{S}^{( i)} = (\partial_{\mb{F}}\widehat\psi)^{( i)} - \bm{S}^{( i)}_p\,,&& i= 0,  1,  2,\\
&\bm{s}^{( i)} = (\partial_{\mb{F}}\widehat\psi)^{({ i})}[\osc{\mb n}] - \bm{s}^{( i)}_{p}\,,&& i= 0, 1\,, \\
&\mu^{({ i})} = (\partial_c \widehat\psi)^{({ i})} + \Omega p^{( i)}\,,&&  i= 0, 1\,. \label{eq:8a}
\end{align}
\end{subequations}
Using \eqref{eq:8}, we obtain from \eqref{eq:3} the following reduced dissipation inequality
\begin{align}
\label{eq:reddissin}
\begin{split}
&-\int_{\mc{P}}\left(\bm{h}^{(\rm 0)}\cdot\grads{{\mu}_{\rm 0}}+h_3^{(\rm 0)}{\mu}_{ 1}+\bm{h}^{(1)}\cdot\grads{{\mu}_{ 1}}\right)\mbox{d}a\ge 0.
\end{split}
\end{align}
A selection criterion for the constitutive equations governing the fluxes $\bm h^{(0)}$, $\bm h^{(1)}$ and $h_3^{(0)}$ can be obtained by requiring consistency with the three-dimensional constitutive law of Darcy type
\begin{align}\label{eq:42}
\mb{h} = -\frac{cD}{\mc{R}T}\nabla\mu\,,
\end{align}
where $D$ is the solvent diffusivity, $\mc{R}$ is the universal gas constant and $T$ is the absolute temperature. Integrating along the thickness in accord with the definitions \eqref{eq:momflux}, and using the representations \eqref{eq:conc} and \eqref{eq:chempot}, we have
\begin{subequations}
\label{eq:redfluxconstplate}
	\begin{align}
	&\bm h^{(0)}= -\frac{D}{\mc{R} T}h\left( c_0\grads \mu_0 + \frac{h^2}{12}c_1\grads\mu_1\right)\,, \\
	&\bm h^{(1)}= -\frac{D}{\mc{R} T}\frac{h^3}{12}\left(c_1 \grads \mu_0+c_0\grads\mu_1\right)\,, \\
	&h_3^{(0)}= -\frac{c_0 D}{\mc{R} T}h\,\mu_1\,.
	\end{align}
\end{subequations}

In view of \eqref{eq:8a}, and consistent with representation \eqref{eq:chempot} of the chemical potential, we write the pressure as a linear function of $\zeta$:
\begin{align}
\label{eq:linearpress}
p(s,\zeta)=p_{\rm 0}(s)+\zeta p_{1}(s)\,.
\end{align}
With this and the definitions \eqref{eq:qs}, the reactive stress resultants \eqref{eq:reactivestress} can be rendered explicitly as
\begin{subequations}
\label{eq:redreactivestress}
\begin{align}
\bm S^{(0)}_p &= h\left(p_{0} + p_{1}\frac{h^2}{12}\deltanu\right) \mb{R}_0 + \frac{h^3}{12}\left(p_1 + p_0\frac{3}{2}\deltanu\right) \mb{R}_1\,, \label{eq:redreactivestress:a}\\
\bm{S}^{(1)}_{p} &= \frac{h^3}{12}\left(\deltanu p_{0} + p_{1}\right) \mb{R}_0 + \frac{h^3}{12}p_{0}\mb{R}_1\,, \\
\bm{S}^{(2)}_{p}  &= \frac{h^3}{12} p_0 \mb{R}_0\,, \\
\bm{s}^{(0)}_{p} &= h p_0 \left(\mb{r}_0 + \frac{h^2}{24}\mb{r}_2 \right) + \frac{h^3}{12} p_1 \mb{r}_1\,, \\
\bm{s}^{(1)}_{p} &= \frac{h^3}{12} ( p_0 \mb{r}_1 + p_1 \mb{r}_0 )\,,
\end{align}
\end{subequations}
where
\begin{align}
&{\mb{R}_0} = \frac{1}{{\osc\Gamma}}\left[(\mb{a}_2\times\mb{d})\otimes\osc{\mb g}_{1} - (\mb{a}_{ 1}\times\mb{d})\otimes\osc{\mb g}_2\right]\,, && {\mb{R}_1} = \frac{1}{{\osc\Gamma}}\left[(\mb{d}_{,2}\times\mb{d})\otimes\osc{\mb g}_{1} - (\mb{d}_{,1}\times\mb{d})\otimes\osc{\mb g}_2\right]\,,
\end{align}
where we have to retained terms up to $O(h^3)$.
Then, recalling  \eqref{eq:linearpress}, we deduce the following relations from \eqref{eq:8a}:
\begin{subequations}\label{eq:13}
	\begin{align}
	&\mu^{(0)}= (\partial_c\widehat\psi)^{({0})} + \Omega h p_{0}\,, \\
	&\mu^{(1)} = (\partial_c\widehat\psi)^{({1})} + \Omega \frac {h^3}{12}p_{1}\,.
 \end{align}
\end{subequations}
A handier constitutive equation can be obtained by recalling that $c(s,\zeta)=c_0(s)+\zeta c_1(s)$ and the Taylor expansion 
\begin{equation*}\partial_c\widehat\psi(\mathbf F,c_0)=\partial_c\widehat\psi(\mathbf F,c_0)+\partial^2_c\widehat\psi(\mathbf F,c_0)\zeta c_1+\frac{\partial^3_c\widehat\psi(\mathbf F,c_0)}2\zeta^2 c_1^2+o(\zeta^2)
\end{equation*}
so that, by substituting in \eqref{eq:13}, and using $\mu^{(0)}= h\mu_{0}$ and $\mu^{(1)}=\frac{h^3}{12}\mu_1$ we obtain
\begin{subequations}\label{eq:15}
\begin{align}
	&\mu_{ 0} = \,\partial_c \widehat\psi(\mathbf F,c_{ 0}) + \frac{h^2}{24}\,\partial^3_c \wh{\psi}(\mathbf F,c_{ 0})c_{ 1}^2  + \Omega p_{ 0}\,, \\
	&\mu_{ 1} = \partial^2_c \widehat\psi(\mathbf F,c_{ 0}) c_{ 1} + \Omega p_{ 1}\,.
\end{align}
\end{subequations}

\subsection{The Flory-Rehner free energy}
We assume that the reference configuration $\mc{B}$ of the polymer gel is attained through a homogeneous swelling from the dry state that produces a spatially--uniform spherical distortion $\mathbf F_\star=\lambda_\star\mathbf I$, with $\lambda_\star\ge 1$. Hence, the deformation gradient with respect to the dry state and the polymer volume fraction in the current configuration are given by, respectively,
\begin{equation}
\label{eq:polymvolfrac}
\mbox{
 $\mathbf F_{\rm d}=\lambda_\star\mathbf F$\qquad  and \qquad $\phi=1/(J_\star\operatorname{det}\mathbf F)$\,,}
\end{equation} 
where $J_\star=\operatorname{det}\mathbf F_\star$ and in the computation of $\phi$ we have assumed that the polymer network is incompressible.
According to Gaussian network theory, the strain energy per \emph{unit dry volume} is (e.g. \cite[Eq. (3.8)]{Doi2009JPSJ})
\[
\widetilde\psi_{\rm s}(\mathbf F_{\rm d})=\frac {G}{2}(|\mathbf F_{\rm d}|^2-3),
\]
where $G$ is the shear modulus of the polymer network. Thus, the function
\begin{equation}\label{eq:12}
\widehat\psi_{\rm s}(\mathbf F)=\frac{1}{J_\star}\widetilde\psi_{\rm s}(\lambda_\star\mathbf F)=\frac{G}{2J_{\star}}(|\lambda_\star\mathbf F|^2-3)
\end{equation}
yields the dependence on $\mathbf F$ of the strain energy per unit \emph{reference volume}. Moreover, for $\mathcal R$ the \emph{universal gas constant}, $T$ the \emph{absolute temperature}, and $\chi$ the \emph{solvent-polymer interaction parameter}, the function (see \cite[Eq. (3.68)]{Doi2009JPSJ})
\begin{equation*}
  \widetilde\psi_{\rm m}(\phi)=\frac{\mathcal RT}{\Omega}\left(\frac 1\phi-1\right)(\log(1-\phi)+\chi\phi)
\end{equation*}
accounts for the mixing energy per unit \emph{dry volume} according to the Flory-Huggins solution theory. Now, given that the amount of solvent per unit dry volume is $J_\star c$, the volume fraction in the current configuration is $\phi=1/(1+\Omega J_\star c)$. Accordingly, given the volume constraint  \eqref{eq:volumeconstr} and  \eqref{eq:polymvolfrac}$_2$, the initial solvent concentration per unit reference volume is $c_\star = (J_\star-1)/\Omega J_\star$. Moreover, 
\begin{equation}\label{eq:11}
  \widehat\psi_{\rm m}(c)=\frac 1 {J_\star}\widetilde\psi_{\rm m}\left(\frac 1 {1+\Omega J_\star c }\right)=\mc{R}Tc\left[\log\left(\frac{\Omega J_{\star} c}{1+\Omega J_{\star} c}\right)+\chi \frac{1}{1+\Omega J_{\star} c} \right]
\end{equation}
is the mixing energy per unit reference volume. Summing up,
\begin{align}\label{eq:1}
	&\widehat\psi(\mb{F},c) = \widehat\psi_{\rm s}(\mb{F})+\widehat\psi_{\rm m}(c)
\end{align}
represents the total Flory-Rehner free energy per unit reference volume. 

In view of \eqref{eq:1} and \eqref{eq:12}, the constitutive prescriptions \eqref{eq:8} take the following specialization:
\begin{subequations}\label{eq:20}
\begin{align}
  	&\bm{S}^{(0)}= h\frac{G}{\lambda_{\star}}\left(\wh{\mb{F}}+\frac{h^2}{24}\left(\mb d\otimes\grads{\deltanu}+{\deltanu}\grads{\mb{d}}\right)\right) - \bm
        {S}^{(0)}_p \,,\label{eq:20a} \\
	&\bm{S}^{(1)}  = \frac{h^3}{12}\frac{G}{\lambda_{\star}}\grads{\mb{d}} - \bm{S}^{(1)}_p\,, \\
	&\bm{S}^{(2)} = \frac{h^3}{12}\frac{G}{\lambda_{\star}}\wh{\mb{F}} - \bm{S}^{(2)}_{p}\,, \\
	&\bm{s}^{(0)} = h\frac{G}{\lambda_{\star}}\mb{d}- \bm{s}^{(0)}_{p}\,, \\
	&\bm{s}^{(1)} = \frac{h^3}{12}\frac{G}{\lambda_{\star}}{\deltanu}\mb{d}- \bm{s}^{(1)}_{p}\,,
\end{align}
\end{subequations}
We next turn to the constitutive equations \eqref{eq:15} which, granted \eqref{eq:1}, become
\begin{subequations}\label{eq:24}
\begin{align}
	&\mu_{ 0} = \,\partial_c \widehat\psi_{\rm m}(c_{ 0}) + \frac{h^2}{24}\,\partial^3_c \widehat\psi_{\rm m}(c_{ 0})c_{ 1}^2  + \Omega p_{ 0}\,, \\
	&\mu_{ 1} = \partial^2_c \widehat\psi_{\rm m}(c_{ 0}) c_{ 1} + \Omega p_{ 1}\,,
\end{align}
\end{subequations}
with
\begin{subequations}
	\begin{align}
	&\partial_c \widehat\psi_{\rm m}(c_{0}) = \mc{R} T \left(\log{\frac{J_{\star} \Omega c_{ 0}}{1+J_{\star} \Omega c_{ 0}}} + \frac{1}{1+ J_{\star} \Omega c_{ 0}}+\frac{\chi}{(1+ J_{\star} \Omega c_{ 0})^2} \right), \\
	&\partial^2_c \widehat\psi_{\rm m}(c_{ 0}) = \frac{\mc{R} T}{c_{ 0} } \frac{ 1+J_{\star} \Omega c_{ 0} (1-2\chi)}{(1+J_{\star} \Omega c_{ 0})^3}, \\
	&\partial^3_c \widehat\psi_{\rm m}(c_{ 0}) = -\frac{\mc{R} T}{c^2_{ 0} } \frac{ 1+4J_{\star} \Omega c_{ 0} + 3 J^2_{\star} \Omega^2 c^2_{ 0} (1-2\chi)}{(1+J_{\star} \Omega c_{ 0})^4}.
	\end{align}
\end{subequations}

\section{Applications}
\label{sec:applications}

In this section, we validate numerically the procedure of dimensional reduction by comparing the results obtained with the plate theory with those obtained with the three-dimensional model with reference to three benchmark problems, solved using the finite element method. The three-dimensional model and the related numerical aspects have been described in \citep{AL2013}. For the reader's sake we provide in Table~\ref{tab:equations} below a summary of the initial-boundary value problem that arises from our theory.

\begin{table}[h]
	\footnotesize
	\centering
\begin{tabular}{|l|l|l|}
\hline
\multirow{4}{*}{\bf Unknowns} & {\bf Primary} & {\bf Secondary}\\
\cline{2-3}
& configuration fields $r$, $\mb d$, $\delta$ & stress resultants $\mb N$, $\mb M$, $\mb q$, $\bm\tau$, $\tau$\\
& concentration fields  $c_0$ and $c_1$ & chemical potentials $\mu_0$, $\mu_1$\\
& pressure fields $p_0$, $p_1$ &  fluxes $\bm h^{(0)}$, $\bm h^{(1)}$, $h_3^{(0)}$\\
\hline
\multirow{4}{*}{\bf Equations}         & {\bf Balance \& constraint}                          & {\bf Constitutive }\\
\cline{2-3}
                                       & balance of forces \eqref{eq:43}                      &  constitutive equations \eqref{eq:18}, \eqref{eq:redreactivestress}, \eqref{eq:20}, for the stress resultants\\
                                       & balance of solvent mass \eqref{eq:44}                & constitutive equations \eqref{eq:redfluxconstplate} for the solvent flux\\
                                       & swelling constraint \eqref{eq:redswellconstrplate}   & constitutive equations \eqref{eq:24} for the chemical potential\\
\hline
{\bf Prescribed fields} & \multicolumn{2}{l|}{
\begin{minipage}{40em}
\null

Bulk load resultants $\mb f^{(0)}$, $\overline{\mb f}^{(1)}$, $f_d^{(2)}$ on $\mathcal M\times\mc{I}$;

Boundary loads $\mb t^{(0)}$, $\overline{\mb t}^{(1)}$, $t_d^{(2)}$ on $\partial_{\rm m}^0\mathcal M\times \mc{I}$;

Constraints on the configuration fields $r$, $\mb d$, $\delta$ on $\partial_{\rm m}^1\mathcal M\times \mc{I}$;

Boundary fluxes $\beta^{(0)}$, $\beta^{(1)}$ on $\partial_{\rm c}^0\mathcal M\times \mc{I}$ and $\overline\beta^{(0)}$, $\overline\beta^{(1)}$ on $\mathcal M \times \mc{I}$;

Constraints on the chemical potential fields $\mu_0$, $\mu_1$ on $\partial_{\rm c}^1\mathcal M\times \mc{I}$;

Initial conditions for $c_0$ and $c_1$ on $\mathcal M\times\{0\}$.

\null
\end{minipage}
}\\
\hline
\end{tabular}
	\caption{Data, unknowns, and governing equations of the boundary-value problem arising from our plate theory. The problem is formulated in a space--time domain $\mathcal M\times\mc{I}$, with $\mathcal M$ a two--dimensional region and $\mc{I}$ a time interval. }
	\label{tab:equations}
\end{table}
In what follows, the gel is supposed to be in equilibrium with a solvent at chemical potential $\mu_{\rm e}$. The condition of chemical equilibrium $\mu_0=\mu_{\rm e}, \mu_1 = 0$ is prescribed through a couple of Lagrange multipliers that enter the balance of solvent mass for $c_0$ and $c_1$ as bulk source terms in the way illustrated in the remark at the end of Section~\ref{sec:balanceeq}.

\subsection{Bending of a cantilever plate}
\begin{figure}[!h]
	\centering
	\includegraphics[scale=0.8]{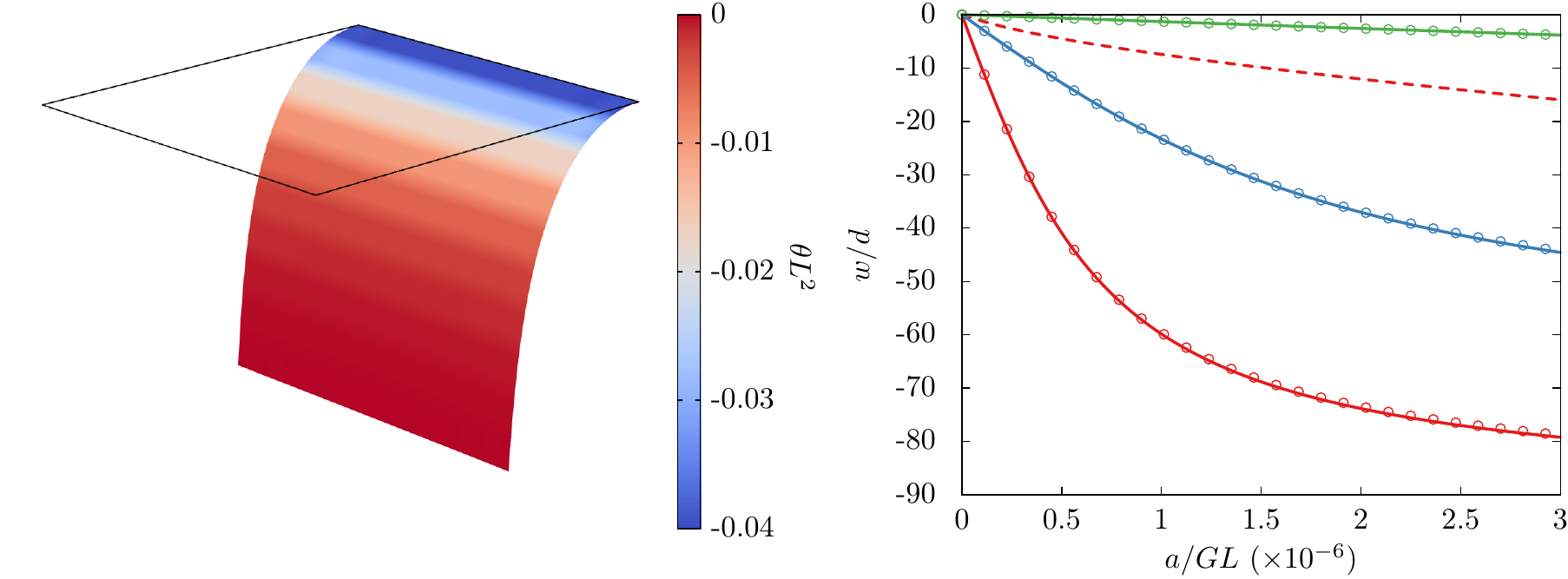}
	\caption{Bending of a square polymer gel plate induced by a vertical edge load with magnitude $a$ applied far from the clamp: comparison between the plate model and the 3D model, for $G\Omega/\mc{R} T = 0.001$, $\lambda_{\star}=1.5$, $\chi = 0.2$. (Left) Deformed equilibrium shape obtained  with the plate model for $h/L = 0.01$ and $a/GL = 3\times 10^{-6}$. The color code represents the dimensionless curvature $\theta L^2$ of the midplane, while the black line represents the undeformed configuration of the midplane. (Right) Dimensionless vertical displacement $w/d$ at the tip of the plate as a function of the magnitude of the edge load, for different values of the thickness-to-edge ratio $h/L$: 0.01 (red), 0.014 (blue), 0.03 (green). Solid lines are the results obtained with the 3D model, while circles are the results obtained with the plate model. The dashed line corresponds to the 6-parameter model.}
	\label{fig:plate_bend_shellvs3D}
\end{figure}

The first problem regards the bending of a cantilever square plate with side length $L$, subject to a vertical edge load $\mb{t}^{(0)}=-a\mb{e}_2, a>0$ applied on the edge opposite to the clamp. The chemical potential $\mu_{\rm e}$ of the external solvent is kept fixed to the value, determined by the parameters $G\Omega/\mc{R} T = 0.001$, $\lambda_{\star}=1.5$, $\chi = 0.2$, which guarantees that the unloaded configuration of the plate is stress-free. As reported in Fig.~\ref{fig:plate_bend_shellvs3D}, the maximum deflection computed using the plate model excellently agrees with that computed using the three-dimensional model, for all the thickness-to-edge ratios $h/L$ and vertical loads considered. In particular, we notice that the model is able to capture large displacements corresponding to the case $h/L=0.01$ at large vertical loads. We also notice that the 6-parameter shell model, where ${\deltanu}\equiv0$, significantly underestimates the deformation of the shell, as observed previously in \citep{braun_nonlinear_1994}.

\subsection{Swelling--induced wrinkling}
\begin{figure}[!h]
	\centering
	\includegraphics[scale=0.8]{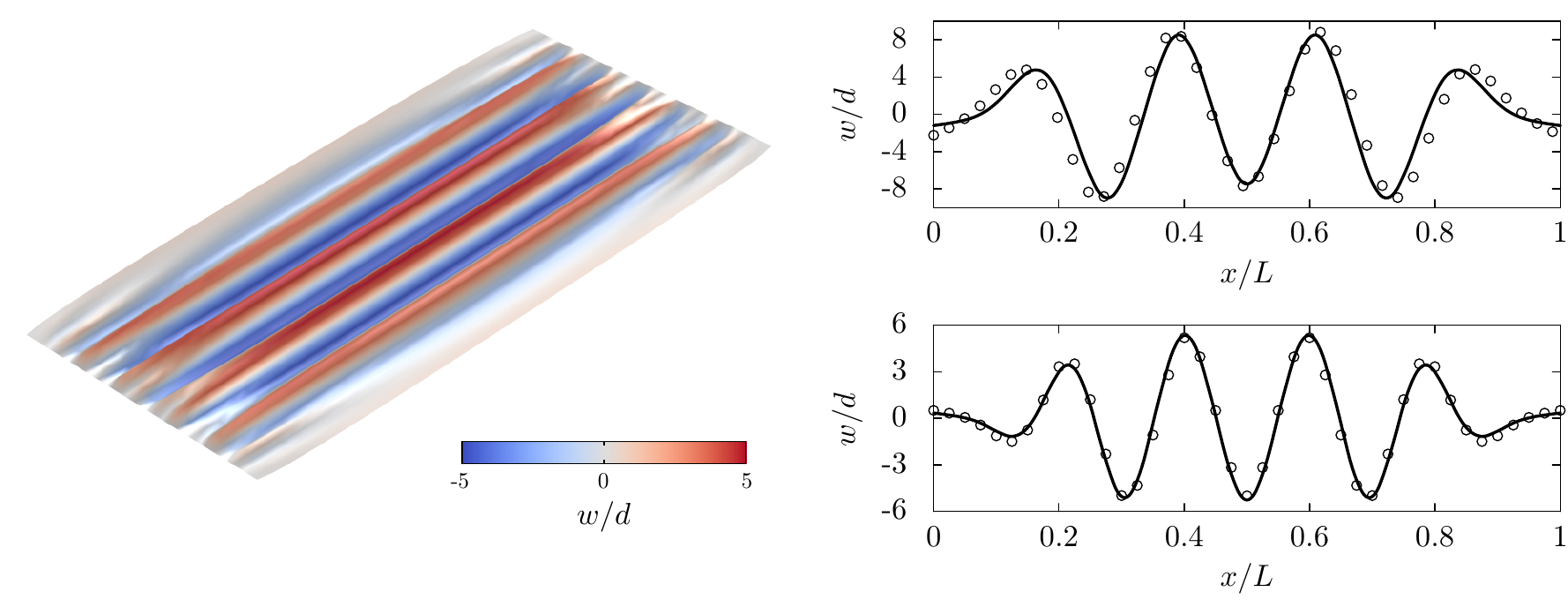}
	\caption{Swelling-induced wrinkling of a square polymer gel plate stretched between clamps: comparison between the plate model and the 3D model, for $G\Omega/\mc{R}T = 0.001$, $\lambda_{\star}=1.5$, $\chi = 0.6$, $h/L = 0.001$ and $\mu_e = 0$. (Left) Deformed equilibrium shape obtained  with the plate model for the applied strain $\varepsilon = 1.0$. The color code represents the dimensionless vertical displacement of the midplane. (Right) Dimensionless vertical displacement $w/d$ along the line $x/L=1/2$ as a function of the dimensionless transverse coordinate $y/L$, for $\varepsilon = 0.5$ (top) and $\varepsilon = 1.0$ (bottom). Solid lines and circles are the results obtained with the 3D model and the plate model, respectively.}
	\label{fig:plate_wrinkle_shellvs3D}
\end{figure}

The second problem regards the swelling-induced wrinkling of a square polymer gel plate pre-stretched between clamps. A similar problem has been studied in \citep{Lucantonio2014a}. First, the clamp-to-clamp distance is increased to fix the nominal strain $\varepsilon = (L'-L)/L$, where $L'$ is the deformed distance, while keeping the chemical potential of the solvent surrounding the plate unchanged, as in the previous example. We consider nominal strains up to $100\%$, where the large-strain constitutive equations for the plate are crucial to accurately evaluate the initial membranal stress field. At the end of the pre-stretch phase, the plate contracts laterally without developing wrinkles. Then, the chemical potential of the external solvent is increased up to $\mu_e=0$, which is higher than the initial chemical potential; as result, the plate absorbs solvent and swells. The constraint imposed by the clamps hampers lateral swelling and thus induces transverse compressive stresses, which trigger the wrinkling instability, as shown in the equilibrium shape of the wrinkled plate depicted in Fig.~\ref{fig:plate_wrinkle_shellvs3D}. Compared to the three-dimensional model, the plate model accurately captures amplitude and wavelength of the equilibrium wrinkling pattern, both decreasing with the nominal strain, in agreement with experimental observations.

\subsection{Polymer gel composite plates}

Inspired by the natural world, where plants adjust their shape by exploiting local changes in swelling, several approaches to shape morphing of polymer gel plates have been proposed. These approaches typically involve the fabrication of polymer gel composite plates \citep{Dickey2016}, either through the embedding of appropriately oriented reinforcing fibers \citep{Erb2013,SydneyGladman2016}, or by introducing a spatial modulation of the cross-linking density of the polymer matrix \citep{Klein2007,Kim2012a,Wu2013}. In-plane stresses arising from non-homogeneous swelling drive the transformation of the initial, flat configuration into complex, three-dimensional shapes. 

Here, we study a problem similar to that presented in \citep{Pezzulla2015}. We consider a circular plate consisting of an inner disk with radius $R_i$ and a circular annulus with external radius $R_e$. The disk and the annulus are made of two polymeric materials with different shear moduli $G_e$ and $G_i$, respectively, obtained by varying the cross-linking density. We fix the geometrical ratios of the structure as $R_i/R_e = 3/10$, $R_e/h = 20$, and the solvent-polymer interaction parameter as $\chi=0.2$. The initial swelling of the stiff and soft materials is determined by the swelling ratio $\lambda_\star \approx 1.8$. Upon increasing the chemical potential of the external solvent up to $\mu_e = 0$, incompatible, differential swelling of the disk and the annulus associated with their non-homogeneous stiffness induces compressive stresses that are relieved by out-of-plane buckling. 

\begin{figure}[!h]
	\centering
	\includegraphics[scale=1.3]{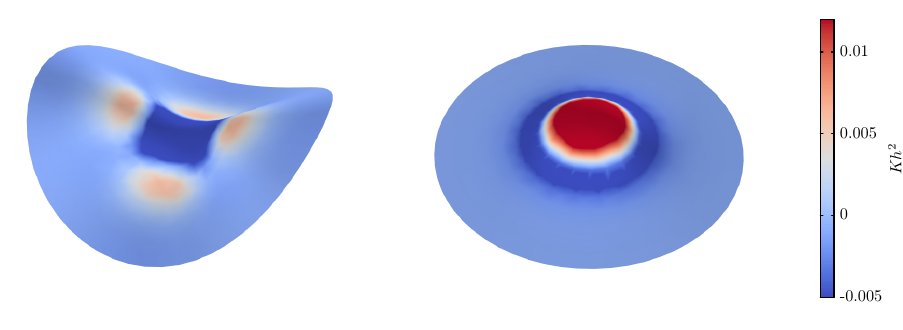}
	\caption{Swelling-induced morphing of saddle-like and dome-like shapes. 
	Deformed equilibrium shapes obtained with the plate model for $G_e/G_i = 3$ (Left) and $G_e/G_i = 1/3$ (Right). The color code represents the dimensionless Gaussian curvature $Kh^2$ of the midplane.}
	\label{fig:composite_plates}
\end{figure}

In the case of stiffer inner disk, the mid-surface of the plate transforms into a saddle-like surface, while in the case of stiffer annulus a dome-like structure is formed, as reported in Fig.~\ref{fig:composite_plates}. As expected, the Gaussian curvature of the deformed mid-surface $K = \det \grads \mb{n}(s)/\det \wh{\mb{F}}$ is mostly negative (positive) in the saddle-like (dome-like) case.

\section{Conclusions}

In conclusion, we have introduced a poroelastic, large-strain plate model that can describe transient, coupled phenomena involving elasticity and solvent migration. The model has been specialized to the case of polymer gels, where the incompressiblity of the polymer matrix and of the solvent induces a kinematic constraint between the volume ratio of the gel and the volume fraction of the solvent. The weak-form plate model has been implemented into a finite element code and its accuracy with respect to the parent three-dimensional model has been demonstrated by several numerical benchmarks. Specifically, we have shown the robustness of the computational model with respect to the analysis of problems involving bifurcations, such as the swelling-induced wrinkling of a pre-stretched membrane and the shape morphing of composite plates. 

We consider the present theory to be relevant in computational mechanics of soft, thin structures as a complement to large-strain, elastic plate models. The computational efficiency with respect to three-dimensional models applied to plate-like structures may be exploited in solving optimization problems associated with the design and patterning of gel-based thin structures. 

\section*{Acknowledgments}
A.L. and A.D.S. acknowledge support from the European Research Council (AdG-340685 -- MicroMotility). A.L. and G.T. acknowledge  support from INdAM-GNFM through the initiative ``Progetto Giovani''.


\bibliography{bibliography-al,bibliography-gt}

\end{document}